\documentclass{article}

\usepackage{comment}
\usepackage{amsthm,amsmath,amsfonts}
\usepackage{graphicx}

\begin{document}
%

\newcommand{\CC}{{\mathbb{C}}}
\newcommand{\NN}{{\mathbb{N}}} 
\newcommand{\ZZ}{{\mathbb{Z}}}
\newcommand{\QQ}{{\mathbb{Q}}}
\newcommand{\gf}{{\sf K}}  
\newcommand{\F}{{\sf F}} 
\newcommand{\diag}{{\rm diag}}
\newcommand{\so}{{O {\;\!\tilde{}}\,}}
\newenvironment{mat}{\left[\!\begin{array}}{\end{array}\!\right]}
\newcommand{\M}{{\sf MM}}
\newcommand{\PM}{{\sf M}}
\newcommand{\m}{{\mathfrak{m}}}
\newcommand{\wrt}{{with respect to}}
\newcommand{\di}{\displaystyle}
\newcommand{\lU}{{\underline U}}
\newcommand{\uU}{{\overline U}}
\newcommand{\nequiv}{\mskip4mu\not\equiv\mskip4mu}
\newcommand{\LinBox}{{\sc LinBox\,}}
\newcommand{\Kry}{{\mathcal{K}}}
\newcommand{\nxn}{{n\times n}}
\newcommand{\nx}[1]{{n\times #1}}
\newcommand{\norm}[1]{\|#1\|}
\newcommand{\divs}{{\mskip3mu|\mskip3mu}}
\newcommand{\ndivs}{\nmid}
\newcommand{\floor}[1]{\lfloor #1 \rfloor}
\newcommand{\litem}{\item[]\noindent\kern-\leftmargin}
\newcommand{\rddots}{\mathinner{
    \mkern2mu\raise1pt\hbox{.}
    \mkern2mu\raise4pt\hbox{.}
    \mkern1mu\raise7pt\vbox{\kern7pt\hbox{.}}
    \mkern1mu}}
\newtheorem{thm}{Theorem}[section]
\newtheorem{lem}[thm]{Lemma}
\newtheorem{cor}[thm]{Corollary}
\newtheorem{ques}[thm]{Question}
\newtheorem{prop}[thm]{Proposition}
\newtheorem{defi}[thm]{Definition}
\newtheorem{rem}[thm]{Remark}
\newtheorem{pty}[thm]{Property}
\newtheorem{con}[thm]{Conjecture}

\newcount\mn \newcount\hr
\def\timeofday{
  \hr=\time \divide\hr 60
  \mn=-\hr \multiply\mn 60 \advance\mn \time
  \ifnum\hr=0
     {12\,:\,\twodigits\mn\,am}
  \else{
     \ifnum\hr<13
        {\number\hr\,:\,\twodigits\mn\,am}
     \else
        {\advance\hr -12 \number\hr\,:\,\twodigits\mn\,pm}
     \fi}
  \fi}
\def\twodigits#1{\ifnum #1<10 0\fi \number#1}
%

\title{Solving Sparse Integer Linear Systems}


\newfont{\adresse}{phvr at 9pt}
\author{
%
 Wayne Eberly,\\
  \adresse{Department of Computer Science, University of Calgary}\\ 
  \adresse{http://pages.cpsc.ucalgary.ca/\~{}eberly}\\  
  \adresse{~~~}\\
 Mark Giesbrecht, Pascal Giorgi\footnote{Author is currently affiliated to LP2A laboratory, University of Perpignan}, Arne Storjohann,\\
  \adresse{David R. Cheriton School of Computer Science, University of Waterloo} \\
  \adresse{http://www.uwaterloo.ca/\,\~{}\{\adresse{mwg,pgiorgi,astorjoh}\}}\\ 
\adresse{~~~}\\
 Gilles Villard\\ 
  \adresse{CNRS, LIP, \'Ecole Normale Sup\'erieure de Lyon}\\
  \adresse{ http://perso.ens-lyon.fr/gilles.villard}\\
}
\date{}

\maketitle

\begin{abstract}
  We propose a new algorithm to solve sparse linear systems of
  equations over the integers.  This algorithm is based on a $p$-adic
  lifting technique combined with the use of block matrices with
  structured blocks. It achieves a sub-cubic complexity in terms of
  machine operations subject to a conjecture on the effectiveness
  of certain sparse projections.
  A \LinBox-based implementation of this algorithm is demonstrated,
  and emphasizes the practical benefits of this new method over the
  previous state of the art.
\end{abstract}






\section{Introduction}

A fundamental problem of linear algebra is to compute the unique
solution of a non-singular system of linear equations.  Aside from its
importance in and of itself, it is key component in many recent proposed
algorithms for other problems involving exact linear systems.  Among
those algorithms are Diophantine system solving
\cite{Giesbrecht:1997:LDE,Mulders:1999:DLSS,Mulders:2004:JSC}, Smith
form computation \cite{egv2000, SaundersWan:2004:issac}, and null-space and
kernel computation \cite{Storjohann:2005:IML}.  In its basic form, the
problem we consider is then to compute the unique rational vector
$A^{-1}b\in \QQ^{n\times 1}$ for a given non-singular matrix
$A\in\ZZ^{n \times n}$ and right hand side $b\in \ZZ^{n \times 1}$.
In this paper we give new and effective techniques for when $A$ is
a sparse integer matrix, which have sub-cubic complexity on
sparse matrices.

A classical and successful approach to solving this problem for dense
integer matrices $A$ was introduced by Dixon in 1982
\cite{Dixon:1982:Pad}, following polynomial case studies from
\cite{MoenckCarter}.  His proposed technique is to compute,
iteratively, a sufficiently accurate $p$-adic approximation $A^{-1}b
\bmod p^k$ of the solution.  The prime $p$ is chosen such that
$\det(A) \nequiv 0 \bmod p$ (see, e.g.,
\cite{Storjohan:2005:HighOrder} for details on the choice of $p$).  Then,
using radix conversion (see e.g. \cite[\S12]{VonzurGathen:1999:MCA})
combined with continued fraction theory \cite[\S10]{HaWr79}, one can
easily reconstruct the rational solution $A^{-1}b$ from $A^{-1}b \bmod
p^k$ (see \cite{Wang:81:issac} for details).

The principal feature of Dixon's technique is the pre-computation of the
matrix $A^{-1} \bmod p$ which leads to a decreased cost of each
lifting step.  This leads to an algorithm with a complexity of
$\so(n^3\log(\norm{A}+\norm{b}))$ bit operations \cite{Dixon:1982:Pad}.  Here
and in the rest of this paper $\norm{\ldots}$ denotes the maximum entry in
absolute value and the $\so$ notation indicates some possibly omitting
logarithmic factor in the variables.

For a given non-singular matrix $A\in\ZZ^{n\times n}$, a right hand
side $b\in\ZZ^{n\times 1}$, and a suitable integer $p$, Dixon's scheme
is the following: 
\begin{list}{$\bullet$}{\topsep=2pt\itemsep=0pt\parsep=2pt}
\item compute $B=A^{-1} \bmod p$;
\item compute $\ell$ $p$-adic digits of the approximation iteratively by
  multiplying $B$ times the right hand side, which is updated according
  to each new digit;
\item use radix conversion and rational number reconstruction
  to recover the solution.  
\end{list} 
The number $\ell$ of lifting steps required to find the exact rational
solution to the system is $\so(n\log(\norm{A}+\norm{b}))$, and one
can easily obtain the announced complexity (each lifting steps
requires a quadratic number of bit operations in the dimension of $A$;
see \cite{Dixon:1982:Pad} for more details).

In this paper we study the case when $A$ is a sparse integer matrix, for
example, when only $\so(n)$ entries are non-zero.  The salient feature
of such a matrix $A$ is that applying $A$, or its transpose, to a
dense vector $c\in\ZZ^{n  \times 1}$ requires only $\so(n\log(\norm{A}+\norm{c}))$
bit operations.

Following techniques proposed by Wiedemann in
\cite{Wiedemann:1986:SSLE}, one can compute a solution of a sparse
linear system over a finite field in $\so(n^2)$ field operations, with
only $\so(n)$ memory.  Kaltofen \& Saunders \cite{Kaltofen:1991:SSLS}
studied the use of Wiedemann's approach, combined with $p$-adic
approximation, for sparse integer linear system.  Nevertheless, this
combination doesn't help to improve the bit complexity compared to
Dixon's algorithm: it still requires $\so(n^3)$ operations in the
worst case.  One of the main reasons is that Wiedemann's technique
requires the computation, for each right hand side, of a new Krylov
subspace, which requires $O(n)$ matrix-vector products by $A\bmod p$.
This implies the requirement of $\Theta(n^2)$ operations modulo $p$
for each lifting step, even for a sparse matrix (and
$\Theta(n(\log\norm{A}+\norm{b}))$ such lifting steps are necessary in
general).  The only advantage then of using Wiedemann's technique is
memory management: only $O(n)$ additional memory is necessary, as
compared to the $O(n^2)$ space needed to store matrix inverse modulo
$p$ explicitly, which may well be dense even for sparse $A$.

The main contribution of this current paper is to provide a new
Krylov-like pre-computation for the $p$-adic algorithm with a sparse matrix
which allows us to improve the bit complexity of linear system
solving.  The main idea is to use block-Krylov method combined with
special block projections to minimize the cost of each lifting step.
The Block Wiedemann algorithm \cite{Coppersmith:1994:SHL,
  Gilles:study, Kaltofen:1995:ACB} would be a natural candidate to
achieve this.  However, the Block Wiedemann method is not obviously
suited to being incorporated into a $p$-adic scheme.  Unlike the
scalar Wiedemann algorithm, wherein the minimal polynomial can be used
for every right-hand side, the Block Wiedemann algorithm needs to use
different linear combinations for each right-hand side.  In
particular, this is due to the special structure of linear
combinations coming from a column of a minimal matrix generating
polynomial (see \cite{Gilles:study,Turner:2002:thesis}) and then be
totally dependent on the right hand side.  

Our new scheme reduces the cost of each lifting step, on a sparse
matrix as above, to $\so(n^{1.5})$ bit operations.  This means the
cost of the entire solver is $\so(n^{2.5}(\log(\norm{A}+\norm{b}))$
bit operations.  The algorithm makes use of the notion of an efficient
sparse projection, for which we currently only offer a
construction which is conjectured to work in all cases.  However, we
do provide some theoretical evidence to support its applicability, and
note its effectiveness in practice.

Most importantly, the new algorithm is shown to offer significant
practical improvement on sparse integer matrices.  The algorithm is
implemented in the \LinBox library \cite{jgd:2002:icms}, a generic C++
library for exact linear algebra.  We compare it against the best
known solvers for integer linear equations, in particular against the
Dixon lifting scheme and Chinese remaindering. We show that in
practice it runs many times faster than previous schemes on matrices
of size greater than $2500\times 2500$ with suffiently high sparsity.
This also demonstrates the effectiveness in practice of so-called
``asymptotically fast'' matrix-polynomial techniques, which employ
fast matrix/polynomial arithmetic.  We provide a detailed
discussion of the implementation, and isolate the performance benefits
and bottlenecks.  A comparison with Maple dense solver emphasizes the
high efficiency of the \LinBox library and the needs of well-designed
sparse solvers as well.



\section{Block projections} 
\label{sec:blockprojection}

The basis for Krylov-type linear algebra algorithms is the notion of a
projection.  In Wiedemann's algorithm, for example, we solve the
ancillary problem of finding the minimal polynomial of a matrix
$A\in\F^{n\times n}$ over a field $\F$ by choosing random
$u\in\F^{1\times n}$ and $v\in\F^{n\times 1}$ and computing the
minimal polynomial of the sequence $uA^iv$ for $i=0..2n-1$ (which is
both easy to compute and with high probability equals the minimal
polynomial of $A$).  As noted in the introduction, our scheme will
ultimately be different, a hybrid Krylov and lifting scheme, but will
still rely on the notion of a structured block projection.

For the remainder of the paper, we adopt the following notation:
\begin{list}{$\bullet$}{\itemsep=0pt\topsep=2pt}
\item $A \in \F^{n \times n}$ be a non-singular matrix,
\item $s$ be a divisor of $n$,  the blocking factor, and
\item $m := n/s$.
\end{list}
Ultimately $\F$ will be $\QQ$ and we will have $A\in\ZZ^\nxn$, but for
now we work in the context of a more general field $\F$.

For a block $v \in \F^{n \times s}$ and $0\leq t \leq m$, define
\[
\Kry(A,v) := \left [ 
  \begin{array}{c|c|c|c} 
    v & Av & \cdots & A^{m-1} v  
  \end{array} \right ] \in \F^{n \times n}.
\]

We call a triple $(R,u,v)\in \F^\nxn \times \F^{s\times n} \times
\F^\nx{s}$ an \emph{efficient block projection} if and
only if
\begin{enumerate}
\item $\Kry(AR,v)$ and $\Kry((AR)^T,u^T)$ are non-singular;
\item $R$ can be applied to a vector with $\so(n)$ operations
  in $\F$;
\item we can compute $vx$, $u^Tx$, $yv$ and $yu^T$ 
for any $x\in\F^{s\times 1}$ and $y\in\F^{1\times n}$, 
with $\so(n)$ operations in $\F$.
\end{enumerate}

In practice we might hope that $R$, $u$ and $v$ in an efficient block
projection are extremely simple, for example $R$ is a diagonal 
matrix and $u$ and $v$ have only $n$ non-zero elements.

\begin{con}
  \label{con:espexists}
  For any non-singular $A\in\F^\nxn$ and $s\divs n$,
  there exists an efficient block projection
  $(R,u,v)\in\F^\nxn\times \F^{s\times n}\times \F^{n\times s}$,
  and it can be constructed quickly.
\end{con}

\subsection{Constructing efficient block projections}

In what follows we present an efficient sparse projection which we
conjecture to be effective for all matrices.  We also present some
supporting evidence (if not proof) for its theoretical effectiveness.  As
we shall see in Section \ref{sec:implementation}, the projection
performs extremely well in practice.  

We focus only on $R$ and $v$, since its existence should imply the
existence of a $u$ of similar structure.

For convenience, assume for now that all elements in $v$ and $R$ are
algebraically independent indeterminates, modulo some imposed
structure.  This is sufficient, since the existence of an efficient
sparse projection with indeterminate entries would imply that a
specialization to an effective sparse projection over $\ZZ_p$ is
guaranteed to work with high probability, for sufficiently large $p$.
We also consider some different possibilities for choosing $R$ and $v$.

\subsubsection{Dense Projections}
The ``usual'' scheme for block matrix algorithms is to choose $R$
diagonal, and $v$ dense.  The argument to show this works has several
steps.  First, $AR$ will have distinct eigenvalues and thus will be
non-derogatory (i.e., its minimal polynomial equals its characteristic
polynomial).  See \cite{cekstv:02:laa}, Lemma 4.1.  Second, for any
non-derogatory matrix $B$ and dense $v$ we have $\Kry(B,v)$
non-singular (see \cite{Ka:95:mathcomp}).  However, a dense $v$ is not
an efficient block projection since condition (2) is not satisfied.

\subsubsection{Structured Projections}
The following projection scheme is the one we use in practice.  Its
effectiveness in implementation is demonstrated in Section
\ref{sec:implementation}.  

Choose $R$ diagonal as before.  Choose
\begin{equation}
  \label{v:sparse}
  \renewcommand{\arraystretch}{2}
  v = \left [ \begin{array}{c|c|c|c} \ast &  & & \\\hline
      & \ast &  & \\\hline
      & & \ddots & \\\hline
      & & & \ast  \end{array} \right ] \in k^{n \times s}
\end{equation}
with each $\ast$ of dimension $m \times 1$.  The intuition behind the
structure of $v$ is twofold.  First, if $s=1$ then $v$ is a dense
column vector, and we know $\Kry(AR,v)$ is non-singular in this case.
Second, since the case $s=1$ requires only $n$ nonzero elements in the
``block'', it seems that $n$ nonzero elements should suffice in the
case $s>1$ also. Third, if $E$ is a diagonal matrix with distinct
eigenvalues then, up to a permutation of the columns, $\Kry(E,v)$ is a
block Vandermonde matrix, each $m \times m$ block defined via $m$
distinct roots, thus non-singular.
In the general case with $s>1$ we ask:
\begin{ques}
  \label{qq} 
  For $R$ diagonal and $v$ as in (\ref{v:sparse}), is $\Kry(AR,v)$
  necessarily nonsingular?
\end{ques}
Our work thus far has not led to a resolution of the question.
However, by focusing on the case $s=2$ we have answered the following
similar question negatively: If $A$ is nonsingular with distinct
eigenvalues and $v$ is as in (\ref{v:sparse}), is $\Kry(A,v)$
necessarily nonsingular?
\begin{lem} 
  If $m=2$ there exists a nonsingular $A$ with distinct eigenvalues
  such that for $v$ as in (\ref{v:sparse}) the matrix $\Kry(A,v)$ is
  singular.
\end{lem}
\begin{proof} We give a counterexample with $n=4$.
  Let 
  \[
  E = \left[ \begin {array}{cccc} 1&0&0&0\\\noalign{\medskip}0&2&0&0
      \\\noalign{\medskip}0&0&3&0\\\noalign{\medskip}0&0&0&4\end
    {array} \right]\hbox{~~and~~} P = \left[ \begin {array}{cccc}
      1&0&0&0\\\noalign{\medskip}0&1&1/4&0
      \\\noalign{\medskip}0&1&1&0\\\noalign{\medskip}0&0&0&1\end
    {array} \right].
  \]
  Define
  \[
  A = 3P^{-1}EP = \left[ \begin {array}{cccc} 3&0&0&0\\\noalign{\medskip}0&5&-1&0
      \\\noalign{\medskip}0&4&10&0\\\noalign{\medskip}0&0&0&12\end {array}
  \right].
  \]
  For the generic block
  \[
  v =  \left[ \begin{array}{cc} a_1& \\ a_2 \\ & b_1 \\ & b_2 
    \end{array} \right]
  \]
  the matrix $\Kry(A,v)$ is singular.  By embedding $A$ into a larger
  block diagonal matrix we can construct a similar counterexample for
  any $n$ and $m=2$.
\end{proof}
Thus, if Question~\ref{qq} has an affirmative answer, then proving it
will necessitate considering the effect of the diagonal preconditioner
$R$ above and beyond the fact that ``$AR$ has distinct eigenvalues''.
For example, are the eigenvalues of $AR$ algebraically independent,
using the fact that entries in $R$ are?  This may already be
sufficient.

\subsubsection{A Positive Result for the Case $s=2$}

For $s=2$ we can prove the effectiveness of our efficient
sparse projection scheme.

Suppose that $A \in \F^{n \times n}$ where $n$ is even and $A$ is
diagonalizable with distinct eigenvalues in an extension of~$\F$. Then
$A = X^{-1} D X \in \F^{n \times n}$ for some diagonal matrix~$D$ with
distinct diagonal entries (in this extension).  Note that the rows
of~$X$ can be permuted (replacing $X$ with $PX$ for some
permutation~$P$),
\[
    A = ((PX)^{-1} (P^{-1}DP) (PX)),
\]
and $P^{-1} D P$ is also a diagonal matrix with distinct
diagonal entries. Consequently we may assume without loss
of generality that the top left $(n/2) \times (n/2)$
submatrix $X_{1, 1}$ of~$X$ is nonsingular. Suppose that
\[
X = \left[ \begin{array}{cc} X_{1, 1}&X_{1, 2}\\
              X_{2, 1}&X_{2, 2}\end{array} \right]
\]
and consider the decomposition
\begin{equation}
  \label{eq:definition_of_Z}
  A = Z^{-1} \widehat{A} Z,
\end{equation}
where
\[
Z = \left[\begin{array}{cc}X_{1, 1}^{-1}&0 \\
            0 & X_{1, 1}^{-1}\end{array}\right]
\quad
X = \left[\begin{array}{cc} I&Z_{1, 2}\\
                            Z_{2, 1}&Z_{2, 2}\end{array}\right]
\]
for $n/2 \times n/2$ matrices $Z_{1, 2}$, $Z_{2, 1}$, and~$Z_{2, 2}$,
and where 
\[
\widehat{A} = \left[\begin{array}{cc} X_{1, 1}^{-1}& 0 \\
             0 &X_{1, 1}^{-1}\end{array} \right] D
              \left[\begin{array}{cc} X_{1, 1} & 0\\
             0&X_{1, 1}\end{array} \right],
\]
so that
\[
\widehat{A} = \left[\begin{array}{cc} A_1& 0\\ 0&A_2 \end{array}\right],
\]
for matrices $A_1$ and~$A_2$. The matrices $A_1$ and~$A_2$ are each
diagonalizable over an extension of~$\F$, since $A$ is, and the
eigenvalues of these matrices are also distinct.

Notice that, for vectors $a, b$ with dimension~$n/2$, and for any
nonnegative integer~$i$,
\[
A^i \left[\begin{array}{c}a\\0\end{array}\right]
 = Z^{-1} \widehat{A}^i
  \left[\begin{array}{c}a\\ Z_{2, 1} a\end{array}\right]
\quad \mbox{and} \quad
A^i \left[\begin{array}{c} 0 \\ b\end{array}\right]
= Z^{-1} \widehat{A}^i
   \left[ \begin{array}{c} Z_{1, 2}b \\
        Z_{2, 2}b\end{array} \right].
\]
Thus, if
\[
x = \left[\begin{array}{cc} a\\ Z_{2, 1} a\end{array}\right]
\quad \mbox{and} \quad
y = \left[\begin{array}{cc} Z_{1, 2}b\\ Z_{2, 2}b\end{array}\right]
\]
then the matrix with columns
\[
a, Aa, A^2 a, \ldots, A^{n/2-1} a,
b, Ab, A^2 b, \ldots, A^{n-2-1} b
\]
is nonsingular if and only if the matrix with columns
\[
x, \widehat{A} x, \widehat{A}^2 x, \ldots, \widehat{A}^{n/2-1} x,
y, \widehat{A} y, \widehat{A}^2 y, \ldots, \widehat{A}^{n/2-1} y
\]
is nonsingular. The latter condition \emph{fails} if and only
if there exist polynomials~$f$ and~$g$, each with degree less
than~$n/2$, such that at least one of these polynomials is nonzero and
\begin{equation}
\label{eq:failure_condition_for_this_construction}
f(\widehat{A}) x + g(\widehat{A}) y = 0.
\end{equation}
To proceed, we should therefore determine a condition on~$A$
ensuring that no such polynomials~$f$ and~$g$ exist for some
choice of~$x$ and~$y$ (that is, for some choice of~$a$ and~$b$).

A suitable condition on~$A$ is easily described: We will require
that the top right submatrix~$Z_{1, 2}$ of~$Z$ is nonsingular.

Now suppose that the entries of the vector~$b$ are uniformly and
randomly chosen from some (sufficiently large) subset of~$\F$,
and suppose that $a = -Z_{1, 2} b$. Notice that at least one of~$f$
and~$g$ is nonzero if and only if at least one of $f$ and~$g-f$ is
nonzero. Furthermore,
\[
f(\widehat{A})(x) + g(\widehat{A})(y)
 = f(\widehat{A})(x+y) + (g-f)(\widehat{A})(y).
\]
It follows by the choice of~$a$ that
\[
x+y = \left[\begin{array}{cc} 0\\
         (Z_{2, 2} - Z_{2, 1} Z_{1, 2})b\end{array}\right].
\]
Since $\widehat{A}$ is block diagonal, the top $n/2$ entries of
$f(\widehat{A})(x+y)$ are nonzero as well for every polynomial~$f$.
Consequently, failure
condition~(\ref{eq:failure_condition_for_this_construction})
can only be satisfied if the top $n/2$ entries of the vector
$(g-f)(\widehat{A})(y)$ are also all zero.

Recall that $g-f$ has degree less than~$n/2$ and that the top left
submatrix of the block diagonal matrix~$\widehat{A}$ is diagonalizable
with $n/2$ distinct eigenvalues. Assuming, as noted above,
that $Z_{1, 2}$ is nonsingular (and recalling that the top half
of the vector~$y$ is $Z_{1, 2}b$), the Schwartz-Zippel lemma is easily
used to show that if~$b$ is randomly chosen as described then, with
high probability, the failure condition can only be satisfied if $g-f = 0$.
That is, it can only be satisfied if $f=g$. 

Observe next that, in this case,
\[
f(\widehat{A})(x) + g(\widehat{A})(y) = f(\widehat{A})(x+y),
\]
and recall that the bottom half of the vector $x+y$ is the
vector $(Z_{2, 2} - Z_{2, 1} Z_{1, 2})b$. The matrix
$Z_{2, 2} - Z_{2, 1} Z_{1, 2}$ is clearly nonsingular (it
is a Schur complement formed from~$Z$) so, once again, the
Schwartz-Zippel lemma can be used to show that if $b$ is randomly chosen as
described above then $f(\widehat{A})(x+y) = 0$ if and only if $f=0$ as
well.

Thus if $Z_{1, 2}$ is nonsingular and $a$
and~$b$ are chosen as described above then, with high probability,
equation~(\ref{eq:failure_condition_for_this_construction}) is
satisfied only if $f=g=0$. There must therefore exist a choice
of~$a$ and~$b$ providing an efficient block projection --- once again,
supposing that $Z_{1, 2}$ is nonsingular.

It remains only to describe a simple and efficient randomization
of~$A$ that achieves this condition with high probability:
Let us replace~$A$ with the matrix
\[
\widetilde{A} = 
\left[\begin{array}{cc}I&tI\\ 0&I\end{array}\right]^{-1} A
\left[\begin{array}{cc}I&tI\\ 0&I\end{array}\right]
=
\left[\begin{array}{cc}I&-tI\\ 0&I\end{array}\right] A
\left[\begin{array}{cc}I&tI\\ 0&I\end{array}\right],
\]
where $t$ is chosen uniformly from a sufficiently large subset
of~$\F$. This has the effect of replacing $Z$ with the matrix
\[
Z \left[ \begin{array}{cc} I&tI\\ 0&I\end{array}\right]
= \left[\begin{array}{cc}I&Z_{1, 2} + tI\\
                 Z_{2, 1}&Z_{2, 2} + t Z_{2, 1}\end{array}\right]
\]
(see, again, (\ref{eq:definition_of_Z})),
effectively replacing $Z_{1, 2}$ with $Z_{1, 2} + tI$. There are
clearly at most~$n/2$ choices of~$t$ for which the latter matrix is
singular.

Finally, note that if $v$ is a vector and $i \ge 0$ then
\[
\widetilde{A}^i v =
\left[\begin{array}{cc} I&-tI\\ 0&I \end{array}\right]
   A^i \left[\begin{array}{cc} I&tI\\ 0&I\end{array}\right] v.
\]
It follows by this and similar observations that this
randomization can be applied without increasing the asymptotic cost of
the algorithm described in this paper.

\emph{Question:} Can the above randomization and proof be generalized
to a similar result for larger~$s$?

\subsection*{Other sparse block projections}
Other possible projections are summarized as follows.

\begin{itemize}

\item \textbf{Iterative Choice Projection.}
  Instead of choosing $v$ all at once, choose the columns of
  $v=[v_1|v_2|\cdots | v_s]$ in succession.  For example, suppose up
  to preconditioning we can assume we are working with a $B \in \F^{n
    \times n}$ that is simple as well as has the property that the
  characteristic polynomial is irreducible.  Then we can choose $v_1$
  to be the first column of $I_n$ to achieve $\Kry(B,v_1)
  \in\F^{n\times m}$ of rank $m$.  Next choose $v_2$ to have two
  nonzero entries, locations chosen randomly until
  $[\Kry(B,v_1)|\Kry(B,v_2)] \in \F^{n \times 2m}$ has rank $2m$, etc.
  This gives a $v$ with $m(m+2)/2$ nonzero entries.

  The point of choosing $v$ column by column is that, while choosing
  all of $v$ sparse may have a very small probability of success, the
  success rate for choosing $v_i$ when $v_1,v_2,\ldots,v_{i-1}$ are
  already chosen may be high enought (e.g., maybe only expected
  $O(\log n))$ choices for $v_i$ before success).

\item \textbf{Toeplitz projections.}
Choose $R$ and/or $v$ to have a Toeplitz structure.

\item \textbf{Vandermonde projections.}
Choose $v$ to have a Vandermonde or a Vandermonde-like structure.

\end{itemize}

\section{Non-singular sparse solver}

In this section we show how to employ a block-Krylov type method
combined with the (conjectured) efficient block projections of Section
\ref{sec:blockprojection} to improve the complexity of evaluating the
inverse modulo $p$ of a sparse matrix.  Applying Dixon's $p$-adic
scheme with such an inverse yields an algorithm with better complexity
than previous methods for sparse matrices, i.e., those with a fast
matrix-vector product.  In particular, we express the cost of our
algorithm in terms of the number of applications of the input matrix
to a vector, plus the number of auxiliary operations.

More precisely, given $A\in\ZZ^\nxn$ and $v\in\ZZ^\nx1$, let $\mu(n)$
be the number of operations in $\ZZ$  to compute
$Av$ or $v^TA$.  Then, assuming Conjecture \ref{con:espexists}, our
algorithm requires \linebreak
$\so(n^{1.5}(\log(\norm{A}+\norm{b}))$
matrix-vector products $w\mapsto Aw$ on vectors $w\in\ZZ^\nx1$
with $\norm{w}=O(1)$, plus $\so(n^{2.5}(\log(\norm{A}+\norm{b}))$
additional bit operations.  

Summarizing this for practical purposes, in the common case of a
matrix $A\in\ZZ^\nxn$ with $\so(n)$ constant-sized non-zero entries,
and $b\in\ZZ^\nx1$ with constant-sized entries, we can
compute $A^{-1}b$ with $\so(n^{2.5})$ bit operations.

We achieve this by first introducing a structured inverse of the
matrix $A_p = A \bmod p$ which links the problem to block-Hankel
matrix theory.  We will assume that we have an efficient block
projection $(R,u,v)\in\ZZ_p^\nxn\times\ZZ_p^{s\times n}\times
\ZZ_p^{n\times s}$ for $A_p$, and let $B=AR\in\ZZ_p^\nxn$.  We thus
assume we can evaluate $Bw$ and $w^TB$, for any $w\in\ZZ_p^\nx1$, with
$\so(\mu(n))$ operations in $\ZZ_p$. The proof of the following lemma
is left to the reader.

\begin{lem}
  \label{lem:bhankel}
  Let $B\in\ZZ_p^{n\times n}$ be non-singular, where $n=ms$ for
  $m,s\in\ZZ_{>0}$.  Let $u \in \ZZ_p^{s\times n}$ and $v\in\ZZ_p^{n\times
    s}$ be efficient block projections such that $V =
  [v|Bv|\cdots|B^{m-1}v]\in\ZZ_p^{n \times n}$ and $U^T =
  [u^T|B^Tu^T|\cdots|(B^T)^{m-1}u^T] \in \ZZ_p^{n \times n}$ are
  non-singular.  The matrix $H=UBV \in \ZZ_p^{n \times n}$ is then a
  block-Hankel matrix, and the inverse for $B$ can be written as
  $B^{-1} = VH^{-1}U$.
\end{lem}

In fact
\begin{equation}
  \label{eq:H}
  H=\begin{pmatrix}
    \alpha_1 & \alpha_2 & \cdots &\alpha_{m} \\
    \alpha_2 & \alpha_3 & \cdots &\alpha_{m+1} \\
    \vdots &            &         &    \\
    \alpha_{m} & \alpha_m & \cdots & \alpha_{2m-1}
  \end{pmatrix}
  \in\ZZ_p^\nxn,
\end{equation}
with $\alpha_i=uB^iv\in\ZZ^{s\times s}$ for $i=1\ldots 2m-1$.  $H$ can
thus be computed with $2m-1$ applications of $B$ to a (block) vector
plus $2m-1$ pre-multiplications by $u$, for a total cost of
$2n\mu(n)+\so(n^2)$ operations in $\ZZ_p$.  For a word-sized prime $p$,
we can find $H$ with $\so(n\mu(n))$ bit operations (where, by
``word-sized'', we mean having a constant number of bits, typically 32
or 64, depending upon the register size of the target machine).

We will need to apply $H^{-1}$ to a number of vectors at each lifting
step and so require that this be done efficiently.  We will do this by
fist representing $H^{-1}$ using the off-diagonal inverse formula of
\cite{Labahn:1990:BlockHankel}:
\begin{align*}
H^{-1}= &
\left(\begin{smallmatrix}
\alpha_{m-1} & \cdots & \alpha_0 \\[-3pt]
\vdots & \kern-7pt\rddots & \\
\alpha_0 & &
\end{smallmatrix}\right)
\left(\begin{smallmatrix}
\beta_{m-1}^* & \cdots & \beta_0^*\\[-3pt]
              & \ddots & \vdots \\
              &        & \beta_{m-1}
\end{smallmatrix}\right) \\
& \hspace*{20pt} -
\left(\begin{smallmatrix}
\beta_{m-2} & \cdots & \beta_0 & 0\\[-3pt]
\vdots & \kern-3pt\rddots & \kern-4pt\rddots & \\[-5pt]
\beta_0  & \kern-3.5pt\rddots & &       \\
0 & & & 
\end{smallmatrix}\right)
\left(\begin{smallmatrix}
\alpha_{m}^* & \cdots & \alpha_1 \\[-5pt]
 & \ddots & \cdots \\
 &        & \alpha_m^* 
\end{smallmatrix}\right)
\end{align*}
where $\alpha_i,\alpha_i^*,\beta_i,\beta_i^*\in\ZZ_p^{s\times s}$.

This representation can be computed using the Sigma Basis algorithm of
Beckermann-Labahn \cite{Labahn:1990:BlockHankel}.
We use the version given in \cite{Giorgi:2003:issac} which ensures
the desired complexity in all cases.
This requires $\so(s^3m)$ operations in $\ZZ_p$ (and will only be done
once during the algorithm, as pre-computation to the lifting steps).

The Toeplitz/Hankel forms of the components in this formula allow to
evaluate $H^{-1}w$ for any $w\in\ZZ_p^\nx1$ with $\so(s^2m)$ or $\so(ns)$
operations in $\ZZ_p$ using an FFT-based polynomial multiplication
(see \cite{CanKal91}).  An alternative to computing the inversion
formula would be to use the generalization of the Levinson-Durbin
algorithm in \cite{Kaltofen:1995:ACB}.

\begin{cor}
  Assume that we have pre-computed $H^{-1}\in\ZZ_p^\nxn$ for a word-sized prime $p$.
  Then, for any $v\in\ZZ_p^\nx1$, we can compute $B^{-1}v\bmod p$ with
  $2(m-1)\mu(n)+\so(n(m+s))$ operations in $\ZZ_p$.
\end{cor}
 
\begin{proof}
  By Lemma \ref{lem:bhankel} we can express the application of
  $B^{-1}$ to a vector by an application of $U$, followed by an
  application of $H^{-1}$ followed by an application of $V$.

  To apply $U$ to a vector $w\in\ZZ_p^{n\times 1}$, we note that
  $$(Uw)^T=[(uw)^T,(uBw)^T,\ldots,(uB^{m-1})^Tw)^t]^T.$$  We can find
  this iteratively, for $i=0,\ldots,m-1$, by computing
  $b_i=B^iw=Bb_{i-1}$ (assume $b_0=w$) and $uB^iw=ub_i$, for
  $i=0..{m-1}$ in sequence.  This requires $(m-1)\mu(n)+\so(mn)$
  operations in $\ZZ_p$.

  To apply $V$ to a vector $y\in\ZZ_p^\nx1$, write
  $y=[y_0|y_1|\cdots|y_{m-1}]^T$, where $y_i\in\ZZ_p^s$.  Then
  \begin{align*}
    Vy &= vy_0+Bvy_1+B^2vy_2+\cdots+B^{m-1}vy_{m-1}\\
    &= vx_0+B\left(vx_1+B\left(vx_1+\cdots
        \left((vx_{m-2}+Bvx_{m-1})
          \cdots\right)\right)\right)
  \end{align*}
  which can be accomplished with $m-1$ applications of $B$ and $m$
  applications of the projection $v$.  This requires
  $(m-1)\mu(n)+\so(mn)$ operations in $\ZZ_p$.
\end{proof}

\subsubsection*{P-adic scheme}

We employ the inverse computation described above in the $p$-adic
lifting algorithm of Dixon \cite{Dixon:1982:Pad}.  We briefly describe
the method here and demonstrate its complexity in our setting.

\begin{list}{}{\topsep=0pt\parsep=0pt}
\litem Input: $A\in\ZZ^\nxn$ non-singular, $b\in\ZZ^\nx1$;
\litem Output: $A^{-1}b\in \QQ^\nx1$
\item[(1)] Choose a prime $p$ such that $\det A\nequiv 0\bmod p$;
\item[(2)] Determine an efficient block projection for $A$:\\
           $R,u,v\in\ZZ^\nxn\times\ZZ_p^{s\times n}\times\ZZ_p^{n\times s}$;
           Let $B=AR$;
\item[(3)] Compute $\alpha_i=uB^iv$ for $i=1\ldots 2m-1$ and define
$H$ as in \eqref{eq:H}. Recall that $B^{-1} = VH^{-1}U$;
\item[(4)] Compute the inverse formula of $H^{-1}$ (see above);
\item[(5)] Let $\ell:=\frac{n}{2}\cdot\left\lceil\log_p(n\norm{A}^2)+\log_p((n-1)\norm{A}^2+\norm{b}^2)\right\rceil)$;\\
           \hphantom{Let } $b_0:=b$;
\item[(6)] For $i$ from 0 to $\ell$  do
\item[(7)] \hspace*{20pt} $x_i:=B^{-1}b_i\bmod p$;
\item[(8)] \hspace*{20pt} $b_{i+1}:=p^{-1}(b_i-Bx_i)$
\item[(9)] Reconstruct $x\in\QQ^\nx1$ from $x_\ell$ using rational reconstruction.
\end{list}

\begin{thm}
  The above $p$-adic scheme solves the linear system $A^{-1}b$ with 
  $\so(n^{1.5}(\log(\norm{A}+\norm{b}))$ matrix-vector products by
  $A\bmod p$ (for a machines-word sized prime $p$) plus
  $\so(n^{2.5}(\log(\norm{A}+\norm{b}))$ additional bit-operations.
\end{thm}
\begin{proof}
  The total cost of the algorithm is
  $\so(n\mu(n)+n^2+n\log(\norm{A}+\norm{b})(m\mu+n(m+s))$.  For the
  optimal choice of $s=\sqrt{n}$ and $m=n/s$, this is easily seen to
  equal the stated cost.  The rational reconstruction in the last step
  is easily accomplished using radix conversion (see, e.g.,
  \cite{VonzurGathen:1999:MCA}) combined with continued fraction
  theory,  in a cost  which is dominated by the other operations
  (see \cite{Wang:81:issac} for details).
\end{proof}


\section{Efficient implementation}
\label{sec:implementation}

An implementation of our algorithm has been done in the \LinBox
library \cite{jgd:2002:icms}.  This is a generic C++ library which
offers both high performance and the flexibility to use highly tuned
libraries for critical components. The use of hybrid dense linear
algebra routines \cite{jgd:2004:issac}, based on fast numerical
routine such as BLAS, is one of the successes of the
library. Introducing blocks to solve integer sparse linear systems is
then an advantage since it allows us to use such fast dense routines.  One
can see in Section \ref{ssec:timing} that this becomes necessary to
achieve high performance, even for sparse matrices.

\subsection{Optimizations}

In order to achieve the announced complexity we need to use
asymptotically fast algorithms, in particular to deal with polynomial
arithmetic.  One of the main concerns is then the computation of the
inverse of the block-Hankel matrix and the matrix-vector products with
the block-Hankel/Toeplitz matrix.

Consider the block-Hankel matrix $H\in\ZZ_p^{n \times n}$ defined by
$2m-1$ blocks of dimension $s$ denoted $\alpha_i$  in equation
\eqref{eq:H}.  Let us denote the matrix power series
\[
H(z)=\alpha_1+\alpha_2z+ \hdots + \alpha_{2m-1}z^{2m-2}.
\]
One can compute the off-diagonal inverse formula of $H$ using
\cite[theorem 3.1]{Labahn:1990:BlockHankel} with the computation of
\begin{list}{$\bullet$}{\itemsep=0pt\parsep=2pt\topsep=2pt}
\item two left sigma bases of $[H(z)^t \, | \, I]^T$ of degrees $2m-2$
  and $2m$, and
\item two right sigma bases of $[H(z) \, | \, I]$ of degrees $2m-2$
  and $2m$.
\end{list}

This computation can be done with $\so(s^3m)$ field operation with the
fast algorithm {\it PM-Basis} of \cite{Giorgi:2003:issac}.  However,
the use of a slower algorithm such as {\it M-Basis} of
\cite{Giorgi:2003:issac} will give a complexity of $O(s^3m^2)$ or
$O(n^2s)$ field operations.  In theory, the latter is not a problem
since the optimal $s$ is equal to $\sqrt{n}$, and thus gives a
complexity of $O(n^{2.5})$ field operations, which still yields the
announced complexity.

In practice, we developed implementations for both algorithms ({\it
  M-Basis} and {\it PM-Basis}), using the efficient dense linear
algebra of \cite{jgd:2004:issac} and an FFT-based polynomial matrix
multiplication.  Nevertheless, due to the special structure of the
series to approximate, the use of a third implementation based on a
modified version of {\it M-Basis}, where only half of the first columns
(or rows) of the basis are computed, allows us to achieve the best
performance. Note that the approximation degrees remain small (less
than $1\,000$).

Another important point in our algorithm is the application of the off
diagonal inverse to a vector $x\in\ZZ_p^\nx1$.  This computation
reduces to polynomial matrix-vector product; $x$ is cut into chunks of
size $s$.  Contrary to the block-Hankel matrix inverse computation, we
really need to use fast polynomial arithmetic to achieve our
complexity.  However, we can avoid the use of FFT-based arithmetic
since the evaluation of $H^{-1}$, which is the dominant cost, can be
done only once at the beginning of the lifting.  Let $t=O(m)$ be the
number of evaluation points.  One can evaluate $H^{-1}$ at $t$ points
using Horner's rules with $O(n^2)$ field operations.

Hence, applying $H^{-1}$ in each lifting step reduces to the
evaluation of a vector $y\in\ZZ_p[x]^{s\times 1}$ of degree $m$ at $t$
points, to computing $t$ matrix-vector product of dimension $s$, and
to interpolating the result. The cost for each application of $H^{-1}$
is then $O(m^2s+ms^2)$ field operations, giving $O(n^{1.5})$ field
operations for the optimal choice of $s=m=\sqrt{n}$.  This cost is deduced
easily from Horner's evaluation and Lagrange's interpolation.

To achieve better performances in practice, we use a Vandermonde matrix
and its inverse to perform the evaluation/interpolation steps.  This
allows us to maintain the announced complexity, and to benefit from the fast
dense linear algebra routine of \LinBox library.

\subsection{Timings}
\label{ssec:timing}

We now compare the performance of our new algorithm against the best
known solvers.  As noted earlier, the previously best known complexity
for algorithms solving integer linear systems is
$\so(n^3\log(||A||+||b||))$ bit operations, independent of their
sparsity.  This can be achieved with several algorithms: Wiedemann's
technique combined with the Chinese remainder algorithm
\cite{Wiedemann:1986:SSLE}, Wiedemann's technique combined with
$p$-adic lifting \cite{Kaltofen:1991:SSLS}, or Dixon's algorithm
\cite{Dixon:1982:Pad}.  All of these algorithms are implemented within
the \LinBox library and we ensure they benefits from the optimized
code and libraries to the greatest extent possible.  In our
comparison, we refer to these algorithms by respectively: {\it
  CRA-Wied}, {\it P-adic-Wied} and {\it Dixon}.  In order to give a
timing reference, we also compare against the dense Maple solver. Note
that algorithm used by Maple 10 has a quartic complexity in matrix
dimension.

In the following, matrices are chosen randomly sparse, with fixed or
variable sparsity, and some non-zero diagonal elements are added in
order to ensure the non-singularity.

\begin{table}[htbp]
  \begin{center}
  \begin{tabular}{|c||r|r|r|r|r|}
    \cline{2-6}
    \multicolumn{1}{c|}{} & \multicolumn{1}{|c|}{\em 400}  & \multicolumn{1}{|c|}{\em 900}  & \multicolumn{1}{|c|}{\em 1600} & \multicolumn{1}{|c|}{\em 2500}  & \multicolumn{1}{|c|}{\em 3600} \\
    \cline{2-6}
    \multicolumn{6}{c}{} \\[-0.1cm]
    \hline
   Maple     & $64.7$s   & $849$s   & $11098$s  & $-$  & $-$   \\
   \hline
   \multicolumn{6}{c}{} \\[-0.25cm]
   \hline
   {CRA-Wied}       & $14.8$s   & $168$s   & $1017$s  & $3857$s  & $11452$s   \\
    \hline
    \multicolumn{6}{c}{} \\[-0.25cm]
    \hline
    { P-adic-Wied} & $10.2$s   & $113$s   &  $693$s  & $2629$s  &  $8034$s  \\
    \hline
    \multicolumn{6}{c}{} \\[-0.25cm]
    \hline
    {Dixon} &  $\bf 0.9$s   &  $\bf 10$s   &   $\bf 42$s  &  $\bf 178$s  &   $429$s \\
    \hline
    \multicolumn{6}{c}{} \\[-0.25cm]
    \hline
    Our algo.  &  $2.4$s   &  $15$s   &   $61$s  &  $175$s  &   $\bf 426$s \\
    \hline
  \end{tabular}
  \caption{Solving sparse integer linear system (10 non-zero elts per row) on a Itanium2, 1.3GHz}\label{tab:all}
\end{center}
\end{table}

First, one can see from Table \ref{tab:all} that even if most of the
algorithms have the same complexity, their performance varies widely.
The P-adic-Wied implementation is a bit faster than CRA-Wied since the
matrix reduction modulo a prime number and the minimal polynomial
computation is done only once, contrary to the $\so(n)$ times needed
by CRA. Another important feature of this table is to show the
efficiency of dense \LinBox's routines compared to sparse
routines. One can notice the improvement by a factor $10$ to $20$ with
Dixon. An important point to note is that $O(n)$ sparse matrix-vector
products is not as fast in practice as one dense matrix-vector product.
Our new algorithm completely benefits from this remark and allows it to
achieve similar performances to Dixon on smaller matrices, and to
outperform it for larger matrices.

\begin{figure}[htbp]
  \begin{center}
    \includegraphics[width=0.6\textwidth,angle=-90]{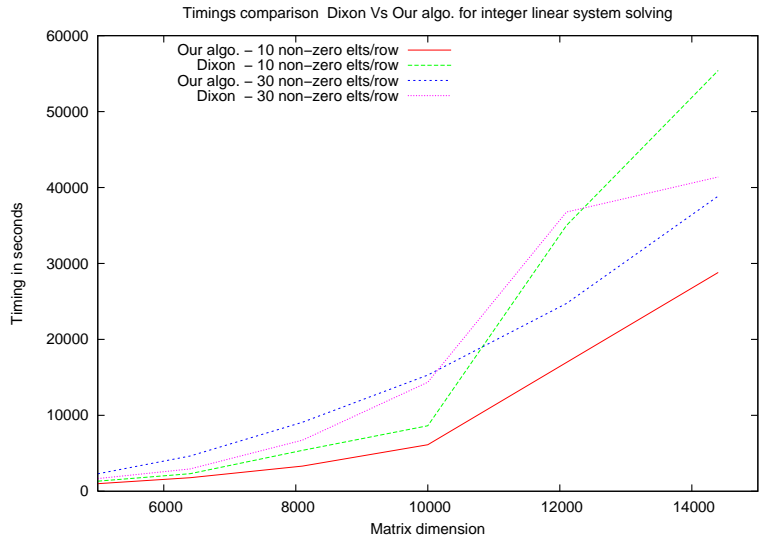}
  \end{center}
  \caption{Comparing our algo. with Dixon's algorithm (fixed sparsity)
    on a Itanium2, 1.3GHz}
  \label{fig:fixsparse}
\end{figure}

In order to emphasize the asymptotic benefit of our new algorithm, we
now compare it on larger matrices with different levels of sparsity.
In Figure \ref{fig:fixsparse}, we study the behaviour of our algorithm
compared to that of Dixon with fixed sparsity (10 and 30 non-zero
elements per rows). The goal is to conserve a fixed exponent in the
complexity of our algorithm.

With $10$ non-zero element per row, our algorithm is always faster
than Dixon's and the gain tends to increase with matrix dimension.
Its not exactly the same behaviour when matrices have $30$ non-zero
element per row.  For small matrices, Dixon still outperforms our
algorithm.  The crossover appears only after dimension $10\,000$.
This phenomenon is explained by the fact that sparse matrix operations
remain too costly compared to dense ones until matrix dimensions
become sufficiently large that the overall asymptotic complexity plays
a more important role.

\begin{figure}[htbp]
  \begin{center}
      \includegraphics[width=0.6\textwidth,angle=-90]{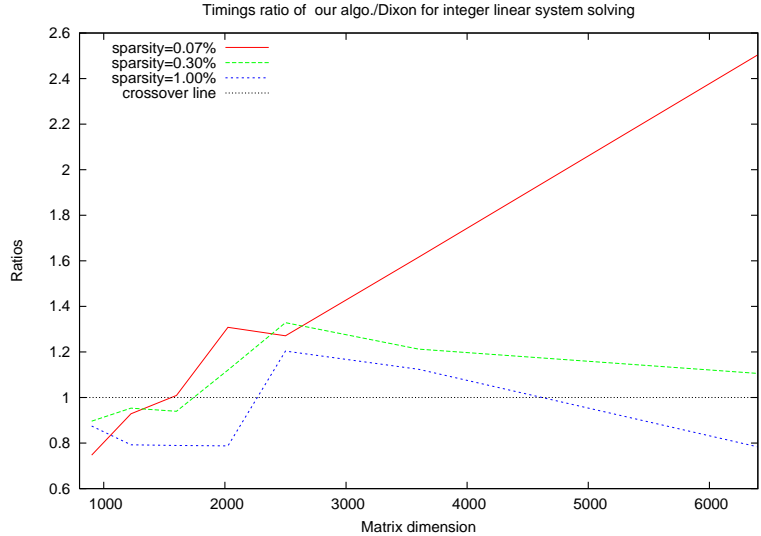}
  \end{center}
  \caption{Gain of our algo. from Dixon's algorithm (variable sparsity) on a Itanium2, 1.3GHz} \label{fig:varsparse}
\end{figure}

This explanation is verified in Figure \ref{fig:varsparse} where
different sparsity percentages are used.  The sparser the matrices
are, the earlier the crossover appears.  For instance, with a sparsity
of $0.07\%$, our algorithm becomes more efficient than Dixon's for
matrices dimension greater than $1600$, while this is only true for
dimension greater than $2500$ with a sparsity of $1\%$.  Another
phenomenon when examining matrices of a fixed percentage density is
emphasized by the Figure \ref{fig:varsparse}.  This is because Dixon's
algorithm again becomes the most efficient, in this case, when the
matrices become large.  This is explained by the variable sparsity
which leads to a variable complexity.  For a given sparsity, the
larger the matrix dimensions the more non-zero entries per row, and
the more costly our algorithm is.  As an example, with $1\%$ of non
zero element, the complexity is doubled from matrix dimension
$n=3\,000$ to $n=6\,000$.  As a consequence, the performances of our
algorithm drop with matrix dimension in this particular case.

\subsection{The practical effect of different blocking factors}

In order to achieve even better performance, one can try to use
different block dimensions rather than the theoretical optimal
$\sqrt{n}$.  The Table \ref{tab:block} studies experimental blocking
factors for matrices of dimension $n=10\,000$ and $n=20\,000$ with a
fixed sparsity of $10$ non-zero elements per rows.

\begin{table}[htbp]
  \begin{center}
    \begin{tabular}{|c||r|r|r|r|r|}
      \cline{2-6}
      \multicolumn{1}{c}{} & \multicolumn{5}{|c|}{\sf n= 10\,000}\\
      \cline{2-6}
      \multicolumn{6}{c}{}\\[-0.2cm]
      \hline
      block size  & \multicolumn{1}{|c|}{\em 80}  & \multicolumn{1}{|c|}{\em 125}  & \multicolumn{1}{|c|}{\em 200} & \multicolumn{1}{|c|}{\em 400} & \multicolumn{1}{|c|}{\em 500} \\
      \hline
      timing &  $7213$s & $5264$s & $4059$s & $\bf 3833$s & $4332$s \\
      \hline 
      \multicolumn{6}{c}{}\\
      \cline{2-6}
      \multicolumn{1}{c}{} & \multicolumn{5}{|c|}{\sf n= 20\,000}\\
      \cline{2-6}
      \multicolumn{6}{c}{}\\[-0.2cm]
      \hline
      block size  & \multicolumn{1}{|c|}{\em 125}  & \multicolumn{1}{|c|}{\em 160}  & \multicolumn{1}{|c|}{\em 200} & \multicolumn{1}{|c|}{\em 500}  & \multicolumn{1}{|c|}{\em 800} \\
      \hline
      timing &  $44720$s & $35967$s & $30854$s & $\bf 28502$s & $37318$s \\
      \hline
    \end{tabular}
    \caption{Blocking factor impact (sparsity= 10 elts per row) on a Itanium2, 1.3GHz}\label{tab:block}
  \end{center}
\end{table}

One notices that the best experimental blocking factors are far from
the optimal theoretical ones (e.g., the best blocking factor is $400$
when $n=10\,000$ whereas theoretically it is $100$).  This behaviour is
not surprising since the larger the blocking factor is, the fewer
sparse matrix operations and the more dense matrix operations are
performed.  As we already noted earlier, operations are performed more
efficiently when they are dense rather than sparse (the cache effect
is of great importance in practice).  However, as shown in Table
\ref{tab:block}, if the block dimensions become too large, the overall
complexity of the algorithm increases and then becomes too important
compared to Dixon's.  A function which should give a good
approximation of the best practical blocking factor would be based on
the practical efficiency of sparse matrix-vector product and dense matrix
operations.  Minimizing the complexity according to this efficiency
would lead to a good candidate blocking factor.  This could be done
automatically at the beginning of the lifting by checking efficiency
of sparse matrix-vector and dense operation for the given matrix.

\section*{Concluding remarks}

We give a new approach to solving sparse linear algebra problems over
the integers by using sparse or structured block projections.  The
algorithm we exhibit works well in practice.  We
demonstrate it on a collection of very large matrices and compare it
against other state-of-the art algorithms.  Its theoretical complexity
is sub-cubic in terms of bit complexity, though it rests still
on a conjecture which is not proven in the general case.  We offer
a rigorous treatment for a small blocking factor (2) and
provide some support for the general construction.

The use of a block-Krylov-like algorithm allows us to link the problem
of solving sparse integer linear systems to polynomial linear algebra,
where we can benefit from both theoretical advances in this field and
from the efficiency of dense linear algebra libraries.  In particular, our
experiments point out a general efficiency issue of sparse linear
algebra: in practice, are (many) sparse operations as fast as
(correspondingly fewer) dense operations?  We have tried to show in
this paper a negative answer to this question.  Therefore, our
approach to providing efficient implementations for sparse linear
algebra problems has been to reduce most of the operations to dense
linear algebra on a smaller scale.  This work demonstrates an 
initial success for this approach (for integer matrices), and it
certainly emphasizes the importance of well-designed (both
theoretically and practically) sparse, symbolic linear algebra algorithms.

\section*{Acknowledgment}
We would like to thank George Labahn for his comments and
assistance on the Hankel matrix inversion algorithms.

\bibliography{eggsv}

\end{document}